  \documentclass{gretsi}

   \usepackage[english,french]{babel}   
  \usepackage{times}			
 
\usepackage{epsfig}
\usepackage{graphicx}
\usepackage{amsmath}
\usepackage{amssymb}
 \usepackage{booktabs}
 \usepackage{multirow}
\usepackage{pifont}
\usepackage{caption}
\usepackage{subcaption}

\usepackage{array}
\usepackage{color}
\usepackage{colortbl}

\usepackage{pifont}
\usepackage{amssymb}
\usepackage{latexsym}
\usepackage{comment}

\title{Deep learning-based image exposure enhancement as a pre-processing for an accurate 3D colon surface reconstruction}

\author{\coord{Ricardo}{Espinosa}{1,3},
        \coord{Axel}{García-Vega}{2},
        \coord{Gilberto}{Ochoa-Ruiz}{2},
        \coord{Dominique}{Lamarque}{4},
        \coord{Christian}{Daul}{3}}

\address{\affil{1}{Universidad Panamericana, Facultad de Ingenier\'ia, Aguascalientes, 20290, M\'exico  
}
         \affil{2}{Escuela de Ingenieria y Ciencias, Tecnologico de Monterrey, 64849 Monterrey, N.L., Mexico 
         }
         \affil{3}{CRAN (UMR 7039), Université de Lorraine and CNRS, 54518 Vandœuvre-l\`es-Nancy cedex, France}
         \affil{4}{H\^opital Ambroise Paré (AP-HP), Boulogne-Billancourt France}
         }

\email{respinosa@up.edu.mx, gilberto.ochoa@tec.mx, christian.daul@univ-lorraine.fr}

\frenchabstract{Cette contribution montre comment un pré-traitement des images peut améliorer la reconstruction 3D du côlon avec de l'apprentissage profond. L'hypothèse est que, comparé à une correction globale de l'illumination, il est plus efficace de corriger les sous- ou sur-expositions locales dans les images. Un aperçu sur la chaîne de traitement qui inclut une correction locale des expositions est d'abord donnée. Ensuite, ce papier quantifie la précision de la reconstruction de la trajectoire de l'endoscope avec et sans correction appropriée des images du côlon.}

\englishabstract{
This contribution shows how an appropriate image pre-processing can improve a deep-learning based 3D reconstruction of colon parts. The assumption is that, rather than global image illumination corrections, local under- and over-exposures should be corrected in colonoscopy. An overview of the pipeline including the image exposure correction and a RNN-SLAM is first given. Then, this paper quantifies the reconstruction accuracy of the endoscope trajectory in the colon with and without appropriate illumination correction.}     

\begin{document}

\maketitle
\section{Introduction}
\vspace*{-3mm}
Colonoscopy is the reference procedure for the visualization  of the inner wall of the large intestine (rectum and colon). It is the only imaging technique that provides natural color and texture tissue information of hollow organs as the colon, stomach and esophagus. Video-sequences are visualized on a screen and analyzed by gastro-enterologists seeking to detect polyps, inflammations, bleeding or cancerous lesions.

However, colonoscopy is a complex examination since the endoscope's trajectory and viewpoint (distance and orientation with respect to the inner tissue) are difficult to control. Besides non-optimal or inappropriate viewpoints for the observation of lesions do not always facilitate the diagnosis, it is difficult to make the camera follow trajectories that ensure a full scan of the tissue with potential lesions. Thus, internal tissue regions can be missed during the visualization, either due to the difficult colonoscope guiding or due to hidden epithelium parts or imaging artifacts such as under or over-exposed frames. 

Moreover, it is difficult to retrieve (i.e., to relocate) a lesion in a given place of a colon segment, either during the examination itself or between two colonoscopies performed in some month intervals. For these reasons, computer vision (CV) methods have been proposed to: i) facilitate lesion (polyp) recognition using supervised classification algorithms, ii) 3D colon part reconstruction to visualize gaps in the internal tissue surfaces, and  iii) image  retrieval algorithms for the determination of the endoscope localization in the organ.

Recently, deep learning (DL) methods have boosted the speed and robustness of the CV approaches in the field of hollow organ endoscopy, notably for the segmentation and recognition of surgical tools \cite{juanca2022}, for the recognition of polyps \cite{9369308}, or for the 3D reconstruction of colon parts using simultaneous localization and mapping (SLAM, \cite{ma2019real}) approaches. 

However, to obtain robust and accurate methods, proper illumination conditions must be ensured and therefore, exposure enhancement pre-processing algorithms in colonoscopy are not only essential for visualization purposes, but are also a prerequisite for the efficiency of the previously mentioned 3D reconstruction algorithms \cite{garciavega2022multiscale}. 
%

One major consequence of uncontrolled colonoscope viewpoints lies in the appearance of multiple artifacts such as under- or over-exposed frames. These frames with insufficient contrast often affect the efficacy  of 3D reconstruction methods, as reported in the literature. This contribution details a method for correcting non-optimal exposures and shows how it improves the camera pose and trajectory estimation yielded by RNN-SLAM (a real-time 3D-reconstruction method) leading to improved estimations colon surface parts. 


The paper is organized as follows. Section 2 gives the motivation of the proposed approach, whereas Section 3 discusses the challenges of the 3D reconstruction of endoscopic data. Section 4 presents the used dataset, as well as the training and exposure correction implementation details. Finally, Section 5 provides colon reconstruction results using the corrected images.


\begin{figure}[!t]
    \centering
    \includegraphics[scale=0.73]{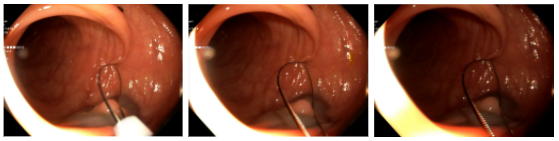}
    \caption{Exposures changes in consecutive video-frames. On the left: an image with an appropriate exposure. In the next images the contrast decreases (under-exposition).}
    \label{sequential}
\end{figure}

\vspace*{-3mm}

\section{Motivation}

The 3D reconstruction of extended hollow organ field of views is an emerging topic based on various CV techniques such as SLAM \cite{ma2019real}, optical flow \cite{TRINH20191} and/or structure from motion (SfM, \cite{PHAN2020107391}). The current trend in colonoscopy has been the integration of DL-based end-to-end reconstruction pipelines. In \cite{ma2019real}, the 3D reconstruction relies on a hybrid scheme combining neural networks and a CV pipeline in which the depth and camera pose are learned using colon data reconstructed with an SfM algorithm. The prediction is used to feed a SLAM-based 3D reconstruction pipeline.


 Many other DL-based reconstruction solutions are inspired by methods initially conceived for autonomous driving. These methods combine Convolutional Neural Networks (CNN) and traditional SLAM approaches for the 3D reconstruction. For instance, CNN-SLAM incorporates CNN-predicted depth maps into the LSD-SLAM framework. Depth maps and pose estimation act as a first step leading to a denser and more accurate uncertainty estimation. Luo et al. \cite{8418760} also replaced the stereo measurements in their Stereo Direct Sparse Odometry (DSO) SLAM with depth values predicted by a CNN. The effectiveness of using a CNN comes from a robust depth prior and gives reasonable depth prediction to assist a SLAM system.

However, for extreme acquisition conditions as in endoscopy, the methods described previously face specific challenges. In colonoscopy or gastroscopy, the scene illumination strongly depends on the endoscope's orientation with respect to the tissue surface, as shown on Fig. \ref{sequential}. Parts of the colon are often under- or over-exposed depending on the surface shape or affected by specular reflections.
%
%
%
Nevertheless, numerous SLAM reconstruction techniques are more or less based on the assumption that the brightness between consecutive images remains almost constant or changes slightly and constantly over the image. This hypothesis ( brightness constancy assumption) is valid to some extent for some applications \cite{Wang_2019_CVPR}, but does not hold in endoscopy where image enhancement is a mandatory step for a robust and clinically usable 3D reconstruction \cite{ma2019real}. 



\vspace*{-3mm}
\section{State-of-the-art}

Endoscopic images are challenging for monocular depth estimation, due to the low-textured surfaces and complex scene illumination.  The authors in \cite{ma2019real} employed an SfM approach to generate sparse depth information using real colonoscopic data and trained in a supervised manner a depth estimation network in real-time solely with these clinical data. Then, they used a a SLAM approach which predicts the camera poses using the colonoscopic images and reconstruct in real-time colon surface parts. This work represents the first successful attempt to reconstruct human colon parts. GAN frameworks and reflection models were used in \cite{Kai-Depth_Estimation} to transfer knowledge learned from synthetic data to real data. This approach enables the visualization of missing tissue regions. However, no image exposure correction step was applied.

Despite the promising results in \cite{ma2019real, Kai-Depth_Estimation}, depth estimation networks can only handle simple photometric changes and may lead to strong depth errors. For complex illumination conditions, the predicted shapes often fail to produce precise surfaces or the colon diameter deviates from its true value. 

Zhang et al. \cite{lightingSLAM} proposed a global  image enhancement technique to improve the 3D reconstruction. In their RNN-SLAM pipeline, an RNN was trained to predict the best gamma value (which models a global intensity change between images) using previous and current frames. The images were enhanced using a histogram equalization and modified by an adaptive gamma correction method.

This previous work has effectively improved the 3D reconstruction of colon parts by exploiting a global exposure correction between images. Nevertheless, using a RNN to only predict the best gamma value for each frame does not allow to correct local under- or over-exposures, and can even produce an over-smoothing of the images. The resulting images can be poorly contrasted (i.e., faded) images.


 The authors in \cite{garciavega2022multiscale} proposed a DL-based image correction called Endo-LMSPEC (Learning Multi-Scale Photo Exposure Correction) which takes into account strong local exposure chan\-ges. This method is able to cope both with under- and over-exposed images in a more local way compared to previous methods \cite{ma2019real, lightingSLAM}. The aim of this paper is to integrate this method into the colon RNN-SLAM reconstruction pipeline.


\section{Materials and Methods}

\subsection{Full pipeline}

\begin{figure*}[ht]
    \centering
    \includegraphics[scale=0.8]{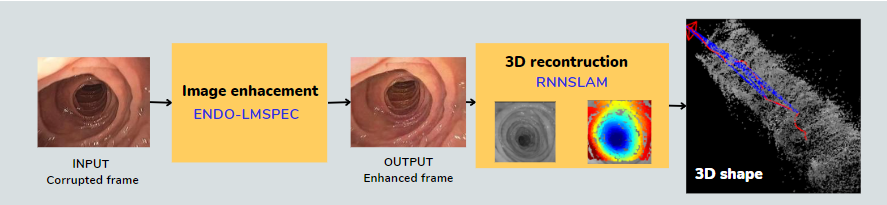}
    \caption{Proposed pipeline including Endo-LSMPEC image enhancement and RNN-SLAM modules.}
    \label{pipeline}
\end{figure*}

The 3D colon reconstruction pipeline which exploits the local image enhancement pre-processing step is given in Fig.~\ref{pipeline}. This pipeline is adapted from RNN-SLAM \cite{ma2019real} which includes tracking, key-frame selection, local windowed optimization, and marginalization modules. The authors associated a RNN with the tracking module and added a fusion module at the pipeline end. This contribution proposes a strategy to combine RNN-SLAM with a deep learning-based exposure enhancement method, namely Endo-LMSPEC. 
%

\subsection{Dataset}

 The RNN was trained using depth maps produced by an SfM technique to deal with  the lack of ground truth (depth maps and camera poses). To do so, the SfM method was used to reconstruct colon parts seen in 60 real colonoscopy video sequences, each containing about 20 thousand frames. Using 200 consecutive frame windows, the SfM technique calculates the corresponding depth maps and camera poses, which are then used for the training (for full details, see the original paper \cite{ma2019real}).

\subsection{Deep learning-based image enhancement}

Rapid and local illumination changes impede reliable point matching and depth predictions which are vital steps to obtain a precise 3D reconstruction. This work uses the deep learning-based image enhancement method called Endo-LMSPEC model which was fed by $n$ small patches $I'_{1},...,I'_{n}$ randomly extracted from the images. Each image patch is used to determine two image pyramids: i) a four-level Gaussian Pyramid (GP) and ii) a four-level Laplacian Pyramid (LP). A LP can be seen as a set of frames with different frequency levels, $LP=\{l_{1}, l_{2}, l_{3}, l_{4}\}$, where $l_{1}$ and $l_{4}$ contain the high- and low-fre\-quen\-cy components, respectively. This LP decomposition is performed to feed four U-Net-like sub-nets in a cascaded configuration with image patches with different levels of detail. Each sub-net is used to extract relevant features from the image and to carry out a reconstruction of each $l_{i}$ input in reverse order; more details can be found in \cite{garciavega2022multiscale}.  The weights of the model trained for the image enhancement phase were only used for the testing set which is composed of 4 real colonoscopy videos. These four image sequences consist of 689 exposure-corrected frames to be reconstructed in the next SLAM steps.

\subsection{Depth \& pose network training}

Figure \ref{pipeline} illustrates how RNN-SLAM utilizes forecasts of the depth and camera position as an initial step for further calculations. The RNN network consists of two parts: a depth estimation network and a camera pose estimation network. The depth estimation network produces a depth map of the same size as the input image, while the camera pose estimation network generates a relative 6-DoF (degree of freedom) camera pose between the current and prior frames. If the camera intrinsic parameters are known, the dense flow field for 2D pixels can be calculated from the current view to the prior view by utilizing the estimated depth map, camera pose, and camera intrinsic parameters. The estimated depth maps and camera poses are then used to generate dense flow fields to warp previous views to the current view through a differentiable geometric module. The training phase was re-implemented on Tensorflow 2.11, using the training set described in section 4.2, in a fully supervised manner over $20$ epochs, using a $0.0002$ learning rate and Adam as optimizer. The model was fed with 10 frames which are grouped in a sliding window fashion, in order to preserve the temporal information between frames (more details in \cite{Wang_2019_CVPR}).
 
\section{Trajectory accuracy and Metrics}

 An open-source tool named EVO \cite{grupp2017evo} was used to quantify the accuracy of the camera trajectory recovery which acts as a quality criterion of the 3D surface reconstruction.  To assess the performance of the local exposure correction, a comparison between the reconstructed endoscope camera trajectories  (with ima\-ge enhancement) and the trajectories obtained with the original RNN-SLAM pipeline (without any enhancement) to ground-truth trajectories must be performed. The latter were generated in an offline way from our testing set of four real colonoscopies. To do so, we used Colmap, a state-of-the-art SfM reconstruction method that incorporates pairwise exhaustive image matching and global bundle adjustment. Colmap is very slow, but it reconstructs trajectories very accurately. For this reason, the ``Colmap trajectories'' can be seen as a ``ground truth'' for the trajectory  accuracy evaluation. 

Two popular metrics were used to compare RNN-SLAM with and without exposure correction, namely the absolute pose error (APE) and the root mean square error (RMSE).  The accuracy of the trajectories provided by the SLAM-methods can be assessed by comparison with the ground truth trajectories. An $APE_i$ value is defined by the $E_i$ values determined with two matrices corresponding to the $i^{th}$ ground truth camera pose $P^{gt}_i$ and to the $i^{th}$ estimated pose $P^{est}_i$ along the camera optical center trajectory $SE$. Pose $P^{gt}_0$ (i.e., $i = 0$) defines the coordinate system in which $SE$ is determined and $trans(E_i)$ refers to the translational components of the relative pose error.

\begin{equation}
E_{i} = (P^{gt}_i)^{-1} P^{est}_i \in SE
\end{equation}

\begin{equation}
APE_{i} = || trans(E_{i})||_{2}
\end{equation}

\begin{equation}
RMSE = \sqrt{\frac{1}{N} \sum\limits_{i=1}^N{APE_{i}^2}}
\end{equation}

 \begin{figure}[th]
    \centering
    \includegraphics[scale=0.8]{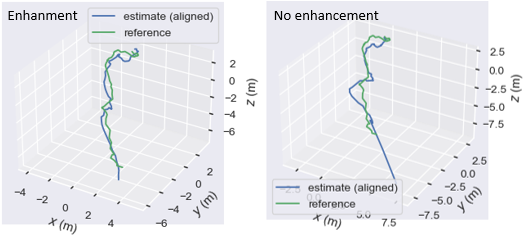}
    \caption{Visual trajectories between the ground truth and the results obtained by RNN-SLAM with and without local exposure corrections.}
    \label{trajectories}
\end{figure}
%
%
 \begin{table}[th]
 \vspace*{-2mm} 
 \normalsize
 \centering
 \caption{Quality metrics. The $\overline{APE}$ and $RMSE$ values are determined by using simultaneously all $AP_i$ values of the four videos, while the $mean$, $std$, $median$ values are computed with the $mean_j$ and $median_j$ of the $AP_i$ values of the four video-sequences ($j\in[1,4]$). The best values are in bold.}
 \label{tab1}
 \scalebox{0.75}{%
 \vspace*{-5mm}
 \begin{tabular}{@{}cccccc@{}}
 \cmidrule(l){1-6}
 RNN-SLAM     & $\overline{APE}$ & $RMSE$   & $std$    & $median$   & $mean$       \\ 
 \cmidrule(l){1-6}
 Without exposure correction  & 1.28            & 0.96            & \textbf{0.13}        & 0.82           & 0.85 \\
With gamma correction \cite{lightingSLAM} & 1.04           & 0.86            & 0.32          & 0.74          & \textbf{0.80} \\
LMSPEC image enhancement\cite{garcia2022novel}   & \textbf{0.97}           &   \textbf{0.66}        &    0.28         &  \textbf{0.68}           &0.85  \\
 \bottomrule
 \end{tabular}%
 } 
 \end{table}


Table \ref{tab1} shows the metric values allowing the assessment of the three approaches, namely RNN-SLAM without image enhancement, RNN-SLAM with gamma correction \cite{lightingSLAM} and RNN-SLAM with LMSPEC (this contribution). The metric values (the lower the better) reflect the proximity of the RNN-SLAM trajectories with their ground truth counterparts (Col\-map-trajectories). It is noticeable that the proposed SLAM method outperforms both RNN-SLAM without enhancement and with gamma correction since the lowest metric values (in bold) are globally in favor of the local exposure correction.
Figure \ref{trajectories} visualizes the trajectories for one colonoscopic video. It can be observed that the proposed SLAM with image correction yield trajectories (blue lines) closer to that of the ground truth.

 \vspace*{-2mm}

\section{Conclusion and future work}
\label{discussion_future_work}
This work shows that an appropriate image pre-pro\-ces\-sing method enables RNN-SLAM based methods to reconstruct more accurately camera trajectories for 3D colonoscopy. Enhancing  under- and over-exposed frames  more locally (in contrast to global gamma correction) has a positive impact on the accuracy of the depth prediction. In future work, an outlier rejection approach will be incorporated in the 3D point cloud generation to improve the meshing and to obtain textured colon parts.

\vspace*{-2mm}

\bibliography{references.bib}{}
\bibliographystyle{IEEEtran}

\end{document}